\documentclass[manuscript,screen,nonacm]{acmart}
\usepackage{url,hyperref,lineno,microtype,subcaption}
\usepackage{braket}

\usepackage{listings}
\usepackage{xcolor}
\usepackage{amsmath, xparse}
\usepackage{soul}

\definecolor{codegreen}{rgb}{0,0.6,0}
\definecolor{codegray}{rgb}{0.5,0.5,0.5}
\definecolor{codepurple}{rgb}{0.58,0,0.82}
\definecolor{backcolour}{rgb}{1,1,1}

\lstdefinestyle{mystyle}{
    backgroundcolor=\color{backcolour},   
    commentstyle=\color{codegreen},
    keywordstyle=\color{magenta},
    numberstyle=\tiny\color{codegray},
    stringstyle=\color{codepurple},
    basicstyle=\ttfamily\footnotesize,
    breakatwhitespace=false,         
    breaklines=true,                 
    captionpos=b,                    
    keepspaces=true,                 
    numbers=left,                    
    numbersep=5pt,                  
    showspaces=false,                
    showstringspaces=false,
    showtabs=false,                  
    tabsize=2
}

\AtBeginDocument{%
  \providecommand\BibTeX{{%
    \normalfont B\kern-0.5em{\scshape i\kern-0.25em b}\kern-0.8em\TeX}}}

\begin{document}

\title{What is Quantum Parallelism, Anyhow? 
}

\author{Stefano Markidis}
\email{markidis@kth.se}
\affiliation{%
  \institution{KTH Royal Institute of Technology}
  \city{Stockholm}
  \country{Sweden}
}

\renewcommand{\shortauthors}{S. Markidis}

\begin{abstract}
Central to the power of quantum computing is the concept of quantum parallelism: quantum systems can explore and process multiple computational paths simultaneously. In this paper, we discuss the elusive nature of quantum parallelism, drawing parallels with classical parallel computing models to elucidate its fundamental characteristics and implications for algorithmic performance. We begin by defining quantum parallelism as arising from the superposition of quantum states, allowing for the exploration of multiple computational paths in parallel. To quantify and visualize quantum parallelism, we introduce the concept of quantum dataflow diagrams, which provide a graphical representation of quantum algorithms and their parallel execution paths. We demonstrate how quantum parallelism can be measured and assessed by analyzing quantum algorithms such as the Quantum Fourier Transform (QFT) and Amplitude Amplification (AA) iterations using quantum dataflow diagrams. Furthermore, we examine the interplay between quantum parallelism and classical parallelism laws, including Amdahl's and Gustafson's laws. While these laws were originally formulated for classical parallel computing systems, we reconsider their applicability in the quantum computing domain. We argue that while classical parallelism laws offer valuable insights, their direct application to quantum computing is limited due to the unique characteristics of quantum parallelism, including the role of destructive interference and the inherent limitations of classical-quantum I/O. Our analysis highlights the need for an increased understanding of quantum parallelism and its implications for algorithm design and performance. 
\end{abstract}


\maketitle

\section{Introduction}
Parallel computing is the simultaneous execution of threads or processes, constituting the fundamental units of computational work. In classical parallel computing, threads/processes are carried out on distinct and often specialized computational units, such as Computing Processing Units (CPU) and Graphics Processing Units (GPU) cores, executing operations serially. Historically, two different forms of parallelism have been categorized, each effectively aligned with specific problem types and formulations: \textit{task} and \textit{data}  parallelism. Task parallelism involves the concurrent execution of potentially diverse tasks on distinct data sets, whereas data parallelism consists of executing the same task on different dataset items. Examples of these two kinds of parallelisms include pipeline parallelism, commonly employed in modern pipelined processors and Operative Systems (OS), as task parallelism, and data-parallel deep-learning training workloads, wherein identical feed-forward and back-propagation operations are performed on different training data.  

Similarly, quantum computing exploits parallelism, performing calculations in the superposition of states, yet in a very distinctive and unique way. Differently from classical silicon- and electronics-based computing systems, consisting of thousands of cores completing serial tasks, a quantum computing core is inherently an elementary parallel computing unit~\cite{deutsch2000machines,lloyd2000ultimate}: calculations can be performed in parallel by unitary transformations acting on a superposition of quantum states~\cite{deutsch1992rapid,marinescu2005promise,chuang1995simple,rieffel2011quantum}. The READ (measurement) and WRITE (initialization) operations remain serial I/O operations as they bridge the quantum to the classical world. In this article, we explore the concept of quantum parallelism, discussing its distinctive fundamentals that render quantum parallelism such an elusive yet fascinating concept.

While the benefits and computational speed-up of parallel computing are nowadays evident and accepted, at the beginning of the first parallel computing systems, an intense diatribe of whether or not any worthwhile benefit in using parallel computing exists, e.g., what is the worth of designing more expensive and complex systems if the returned speed-up or improvement is minimal? At the basis of this discussion, parallelism and its relation to speed-up must be understood. Two influential laws, Amdahl's~\cite{amdahl1967validity} and Gustafson's~\cite{gustafson1988reevaluating,shi1996reevaluating}, emerged from this controversial debate on parallelism's usefulness. The most famous Amdahl's law is summarized by the consideration that the parallel speed-up is fundamentally plagued by the algorithm serial part that asymptotically will lead to a diminishing speed-up performance. Amdahl’s law is an example of the first critical reflection on parallel computing advantage. In a more positive view supporting classical parallelism, Gustafson’s law responds, pointing out that an increased computational workload for each parallel task can solve larger problems and improve the speed-up. 

Today, the scientific community is intensifying discussions on the advantages or the practicality of quantum computing concerning classical computing and whether, in practice, the reach of quantum advantage is achieved, will be achieved, or - in an overly pessimistic position - will ever be achieved~\cite{babbush2021focus,beverland2022assessing,hoefler2023disentangling,fedorov2022quantum,schuld2022quantum}. The classical-quantum I/O bottleneck to input quantum data and need for super-quadratic speed-up are the major concerns~\cite{hoefler2023disentangling}. This work aims to contribute further to this discussion by analyzing the nature of quantum parallelism and how it is used in its application, reconsidering basic classical parallelism strategies and laws, and reformulating them in the quantum context. The basic fundamental research question is: \textit{what are the insights of applying classical parallelism concepts and laws to quantum computing in terms of performance, speed-up, and advantage?} 

To answer this question, we first discuss the nature of quantum parallelism. Using categories and terminology typical of classical computing, we then reconsider quantum parallelism from the classical parallelism perspective and map these concepts to quantum computing. Throughout the article, we rely on using quantum circuit and gate abstractions and pure state descriptions in an ideal set-up (no noise). This conceptual framework allows us to avoid low-level quantum computing hardware concepts, such as pulses and quantum gate implementation. To identify the limits of quantum parallelism, such as the serial part in the classical parallelism laws, we borrow a technique from classical computing to quantify the amount of parallelism. This technique relies on expressing an algorithm and its computation as a graph and then identifying the graph's critical path length~\cite{leiserson2008survive}. We use quantum dataflow diagrams~\cite{vandriessche2013highly} to represent the quantum state superposition evolution. To show the limitations of basic quantum parallelism, we show two exemplary cases of quantum algorithms, the Amplitude Amplification (AA)~\cite{grover1996fast}  and Quantum Fourier Transform (QFT)~\cite{camps2021quantum}  primitives. We show that quantum algorithms exhibit different amounts of quantum parallelism during the execution, depending on the problem input and the amount of interference. We finalize the paper by analyzing the validity of Amdahl's and Gustafson's laws in light of the results of this work. The contributions of this work are the following:
\begin{itemize}
\item We provide an analysis and summary of the fundamental concepts of quantum parallelism. By discussing the fundamental differences between classical parallelism and its quantum counterpart, we offer insights into their distinct usages in applications.
\item We introduce a methodology for characterizing and quantifying quantum parallelism in applications. Leveraging graph-based representations akin to classical dataflow diagrams, we enable precise quantification of parallelism. This methodology is applied to characterize the quantum parallelism inherent in the Quantum Fourier Transform (QFT) and Amplitude Amplification (AA) algorithms.
\item We discuss extensions of the classical concepts of Amdahl's and Gustafson's laws to quantum computing.  
\end{itemize}

\section{The Concept of Quantum Parallelism}
Before delving into the discussion of quantum parallelism in classical terms, it is essential to understand its fundamental nature and how it differs from classical parallelism. This understanding will help us grasp quantum parallelism's usage and potential advantages. The first pioneering work on establishing the foundation of quantum computing by Paul Benioff and Richard Feynman \cite{benioff1980computer,benioff1982quantum,feynman1982simulating,feynman1985quantum} was deeply involved in establishing computing as a physical process, discussing reversibility and dissipation of quantum gates, and exponential growth of resources and did not explicitly mention quantum parallelism. Instead, to our knowledge, the concept of quantum parallelism was first introduced by David Deutsch in 1985 in a seminal paper~\cite{deutsch1985quantum} and further developed by David Deutsch, Richard Jozsa, and Artur Ekert~\cite{jozsa1991characterizing,ekert1996quantum,ekert1998quantum,deutsch1992rapid}.  

\begin{figure}[t]
  \centering
  \includegraphics[width=0.6\columnwidth]{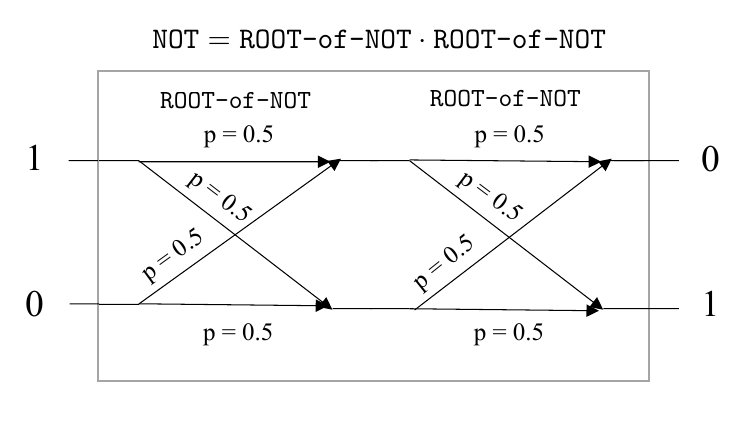}
  \caption{A classical deterministic NOT operation can be decomposed in two quantum \texttt{\footnotesize{ROOT-of-NOT}} operations, exploiting quantum parallelism. $p$ denotes the probability to transition from state 0 to 1 and vice versa. Note that quantum probability addition (instead of classical probability addition) results in a deterministic output. The computation is quantum parallel as it allows for a superposition of states 0 and 1 as execution paths.}
  \label{root}
\end{figure}

Since its inception, quantum parallelism has been the capability of computing systems to perform parallel computation, exploiting the quantum systems' capability of being in a superposition of states. The first articles on the fundamentals of nature quantum parallelism highlight two important points. First, that \ul{computing, intended as a physical process, is inherently quantum parallel}. To convince ourselves, we can go back to the example of implementing a classical NOT on a classical bit as a sequence of two \texttt{\footnotesize{ROOT-of-NOT}} (see Fig.~\ref{root}), as discussed in Refs.~\cite{deutsch2000machines,deutsch2012beyond}. 
In fact, this classical deterministic NOT operation on a classical bit can be implemented using quantum parallelism with two quantum \texttt{\footnotesize{ROOT-of-NOT}} operations (half-NOT operations) and produce a classical output. Even the most basic example of one qubit and NOT operation can exploit parallelism when framed in a quantum computing formulation that is a more fundamental, powerful, and comprehensive theory of classical computing. 

A second important point to note, especially for readers from High-Performance Computing (HPC), is that \ul{quantum parallelism, in its simplest form, does not directly lead to computational speed-up}. Rather, it is a fundamental feature of quantum computing systems. The true source of speed-up lies in the inherent \emph{quantumness} of superposition, which manifests in several critical aspects, summarized as follows:

\noindent \textbf{1. Quantum parallelism follows the principles of interference}, which can either be destructive or constructive. Interference phenomena are rooted in using the squared amplitude of complex numbers to denote probability in quantum computing. A qubit, the fundamental unit of information in two-state quantum systems, can exist in one of two states upon measurement: $\ket{0}$ or $\ket{1}$, corresponding to the standard or computational basis states. Here, $\ket{0}$ represents classical bit 0 and $\ket{1}$ represents classical bit 1. A qubit $\psi$ can also exist in a superposition state, expressed as $\psi = \alpha \ket{0} + \beta \ket{1}$, where $\alpha$ and $\beta$ are complex numbers. These complex numbers can be expressed either in Cartesian form, such as $a + bi$ (where $a$ and $b$ are the real and imaginary components, respectively), or in polar form, $\rho \exp(i \phi)$, where $\rho$ is the magnitude and $\phi$ is an angle. In quantum mechanics, the probabilities of measuring the quantum states $\ket{0}$ or $\ket{1}$ are given by $\left|\alpha\right|^2$ and $\left|\beta\right|^2$, respectively. When a transformation results in a qubit with $\alpha \neq 0$ and $\beta \neq 0$, quantum parallelism is utilized with two quantum threads running simultaneously. The use of complex numbers influences how these parallel threads interact: \ul{probabilities of quantum threads do not obey the additive rule of classical probability laws}; instead, they can cancel out, decreasing the probabilities, via destructive interference. An example of this is evident in the second \texttt{\footnotesize{ROOT-of-NOT}} transformation in Fig.~\ref{root}, where parallelism reduces from two threads to one. Quantum algorithms designed to detect solutions among a limited number of answers must exploit destructive interference to identify the correct answer.

There exists a connection between quantum computing and electromagnetics and antenna theories. It is unsurprising that some quantum algorithms draw inspiration from techniques used in beamforming with antenna arrays~\cite{lloyd1999quantum}: manipulating phases and amplitudes to direct electromagnetic beams in specific directions. Similarly, quantum computing can manipulate the phases and amplitudes of quantum states to achieve the correct solutions through interference. The important point lies in the fact that \ul{quantum parallelism is essentially an interference pattern resulting from the interaction of quantum states}: each quantum state contributes to a collective interference pattern analogous to the combining of signals in antenna arrays. However, unlike classical parallelism, where multiple independent processes run concurrently, quantum parallelism is characterized by a complex web of interference. The notion of reduced quantum parallelism highlights the subtle balance between constructive and destructive interference in quantum algorithms. While constructive interference amplifies certain computational pathways, destructive interference selectively suppresses others, ultimately guiding the system toward the correct solutions. Thus, optimizing interference within quantum algorithms emerges as a critical optimization strategy akin to fine-tuning the amplitude and phase relationships in beamforming applications~\cite{cleve1998quantum}.
\begin{figure}[t]
\centering
\includegraphics[width=0.6\columnwidth]{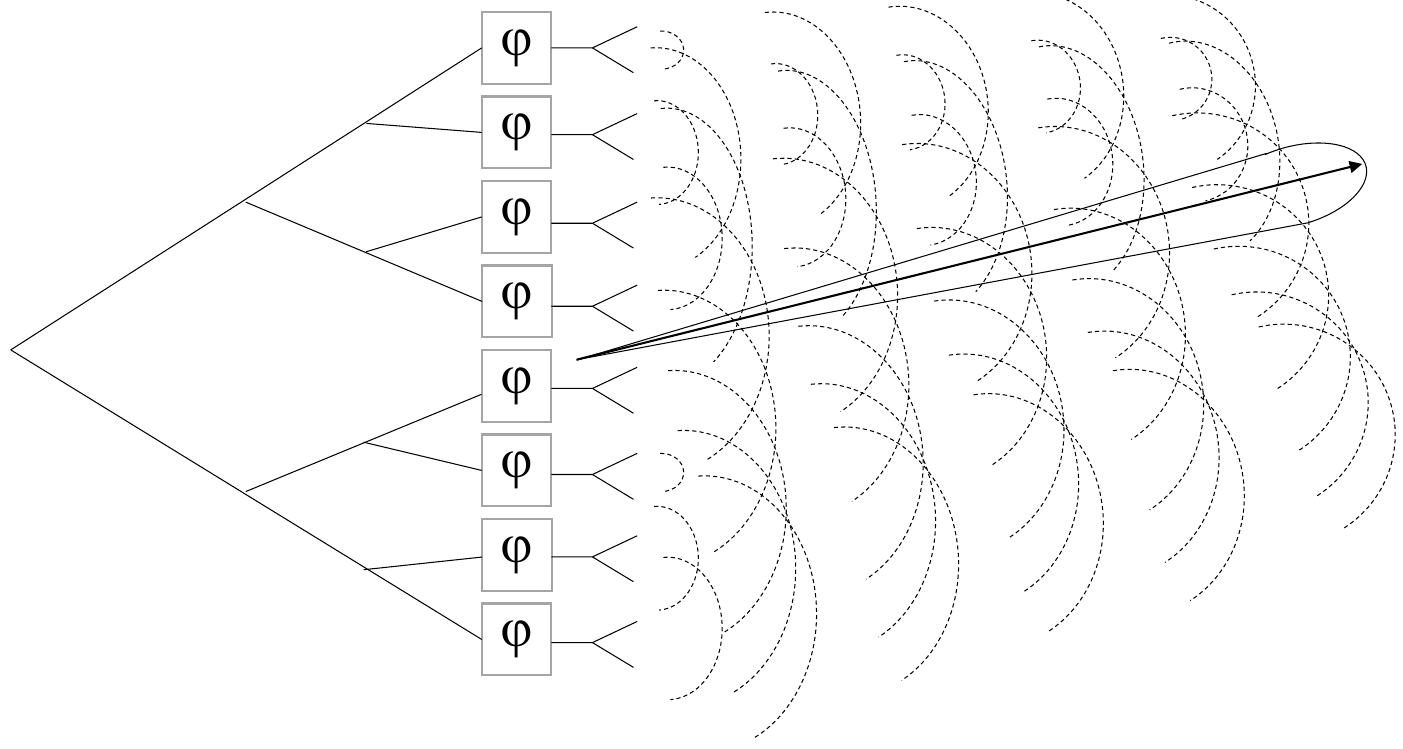}
\caption{Some of the original quantum algorithms were inspired by antenna arrays, where several antennae emit electromagnetic waves with different phases and amplitudes to create directed single or multiple beams. Parallelism arises from utilizing the signal from different antennas. Similarly to quantum computing, beamforming employs parallelism and interference phenomena to cancel wave propagation in undesired directions.}
\label{antennas}
\end{figure}

We might then ask: what is the main difference between classical and quantum interference? The difference lies within the nature of quantum mechanics, where we leverage the complementary nature—also known as \textit{duality}—of particles capable of exhibiting both wave-like and particle-like behavior~\cite{kendon2005complementarity}. This unique property of quantum systems is not exploited in classical systems. In quantum physics, particles such as electrons and photons can manifest as waves under certain conditions, leading to interference phenomena. Even larger entities like atomic nuclei and molecules exhibit wave-like behavior and interference and can be exploited in quantum computing systems to perform computations. 

\noindent \textbf{2. Quantum parallelism exploits the composition of elementary probabilistic systems.} As highlighted in one of the seminal articles on quantum computing by Richard Feynman~\cite{feynman1982simulating}, constructing quantum computing systems involves composing probabilistic systems, leading to an exponential increase in the quantum states of the composite system. Consider two probabilistic systems that do not need to be quantum: $A$ with two possible states ($[A_0, A_1]$) and an associated transition or adjacency matrix $M_A$, and $B$ with three possible states ($[B_0, B_1, B_2]$) and an associated transition or adjacency matrix $M_B$~\cite{yanofsky2008quantum}. The combined states are obtained by taking the tensor product of the two systems' states: $[A_0 B_0, A_0 B_1, A_0 B_2, A_1 B_0, A_1 B_1, A_1 B_2]$. Here, $A_0 B_0$ represents the probability of concurrently observing $A$ in state $A_0$ and $B$ in state $B_0$, and so forth for other elements of the tensor product. Similarly, the combined transformation associated with the assembled state is obtained by taking the tensor product of the adjacency matrices $M_A \otimes  M_B$. The tensor product allows for the assembly of probabilities across different systems. When employing two-state systems, such as qubits, the assembly of $N$ elementary systems results in a combined state with a size of $2^N$, and transitions between different states are expressed by a $2^N \times 2^N$ matrix system. If we used only one classical non-probabilistic system,  such as an antenna array, we would require $2^N$ antennae. Instead, we can leverage the assembly properties of probabilistic systems to scale available resources exponentially. This aspect of quantum parallelism is intriguing: the quantum parallelism available in a multi-qubit system scales exponentially with the number of qubits, with a 300-qubit system offering parallelism greater than the number of particles in the entire universe.

\begin{figure}[t]
 \centering
 \includegraphics[width=0.4\columnwidth]{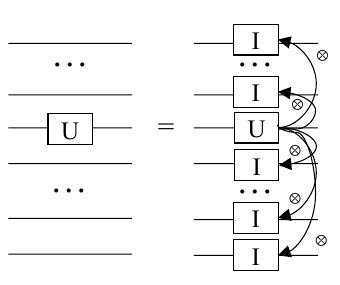}
\caption{A local gate operation $U$ on a qubit is effectively carried out across all  $2^N$ quantum states. While $U$ corresponds to a $2 \times 2$ matrix operation in a single-qubit system, it extends to a  $2^N \times 2^N$ matrix operation in a multi-qubit system. This principle is often referred to as the \textit{principle of local operations}.}
 \label{tensor}
\end{figure}

An important insight into the power of quantum computing arises from the so-called \textit{principle of local operations}, as discussed in Refs.~\cite{deutsch2000concepts,jozsa1998quantum,ekert1998quantum}, wherein a local transformation to a qubit is carried out throughout the entire system. Specifically, the operation $U$ on a certain qubit in a multi-qubit system is expressed as $I \otimes ... \otimes I \otimes U \otimes I .... \otimes I \otimes I$, where $I$ is the identity matrix. See Fig.~\ref{tensor} for a visual representation. This single operation, applied across the entire system, would necessitate the product of a $2^N \times 2^N$ matrix with a vector of size $2^N$ on a classical computer.
While the exponential expansion of space arises from the probabilistic composition of single states, it is crucial to note a significant distinction between classical probabilistic systems and quantum computing systems. In classical probabilistic systems, all events, such as transitions to different states, are mutually exclusive, whereas \ul{quantum computing systems simultaneously explore and process multiple computational paths}. The assembly of elemental quantum computing systems can thus be viewed as a form of superposition creating quantum parallelism. Another critical difference stems from using complex numbers rather than real numbers to express probabilities. In quantum computing, transitions are described by unitary transformations, which, along with reversibility, ensure that the total probability sums to one (unitary transformations \textit{preserve the geometry} of the space in which they operate).

Finally, another significant consequence of assembling basic quantum computing units is evident when considering classical interconnects, such as current HPC networks. Without leveraging probabilistic computing, they cannot exploit exponential parallelism growth. Instead, a quantum interconnect is needed to increase the system's available quantum parallelism exponentially.

\noindent \textbf{3. Entanglement is crucial for quantum parallelism to explore states and threads that are otherwise inaccessible.} As discussed previously, the composition of single-qubit gates generates a state vector that can be produced by composing single-qubit states via the tensor product, thereby excluding states that cannot be represented in this manner. These excluded states, known as \textit{entangled} states, necessitate an entanglement operation for their creation. The fundamental operations that provide entanglement are the controlled operations.

An important aspect concerning entangled states and classical HPC systems, particularly in their role as quantum computer simulators, is that quantum entangled states pose challenges for representation on classical computer systems, requiring exponential resource growth. While non-entangled states, or \textit{separable} states, can be encoded in basic single qubits, allowing for the reconstruction of the full state vector via the tensor product, this encoding method fails when representing an entangled quantum state. This challenge is also evident in tensor-network-based quantum computer simulators, which can efficiently compress quantum states with low entanglement but struggle with expressing quantum systems with high entanglement~\cite{jozsa2006simulation}.
\\

\noindent \textbf{Multi-Core Quantum Computing.}
A multi-qubit system can be likened to a quantum single-core capable of providing $2^N$ quantum parallelism. However, the effort to pack and scale more qubits into a monolithic quantum core presents technological challenges. These challenges include dense wiring requirements for control systems, increased crosstalk between qubits, and the necessity for increased qubit connectivity to enable multi-qubit gate operations, among others. To address the obstacle of scaling the number of qubits, a viable solution is to adopt modular or multi-core quantum architectures~\cite{jnane2022multicore}, wherein different quantum single-cores are interconnected via a quantum network. For example, to realize a 200-qubit Quantum Processing Unit (QPU), ten 20-qubit quantum cores could be combined, circumventing the complexity associated with implementing a single 200-qubit core, which is technologically challenging.

The underlying mechanism of quantum networks and communication relies fundamentally on entanglement. While entanglement in quantum parallelism enables the exploration of quantum state configurations inaccessible via single-gate quantum transformations, in quantum networks, entanglement -- in particular, Einstein--Podolsky--Rosen (EPR) pairs and quantum teleportation -- enables intimately correlated qubits and allows communication to sidestep the constraints of the non-cloning theorem. Analogous to classical computing, inter-core communication in quantum systems is time-consuming and necessitates additional resources, such as dedicated quantum state communication cores (distinct from computational quantum cores) and qubit highways~\cite{zhang2023compilation}. Moreover, quantum networks are highly sensitive to latency, given the limited lifetime of qubits. Consequently, performance models have been developed to evaluate quantum communication costs across the network~\cite{rodrigo2021modelling}, and message-passing programming models~\cite{haner2021distributed, nguyen2023reference}, including collective communication primitives, have been devised to program quantum distributed algorithms. While our discussion primarily focuses on quantum single-core parallelism, the framework presented in this paper can be extended to encompass quantum distributed applications.

\section{The Role of Quantum Parallelism in Quantum Applications}
\begin{figure}[t]
  \centering
  \includegraphics[width=\columnwidth]{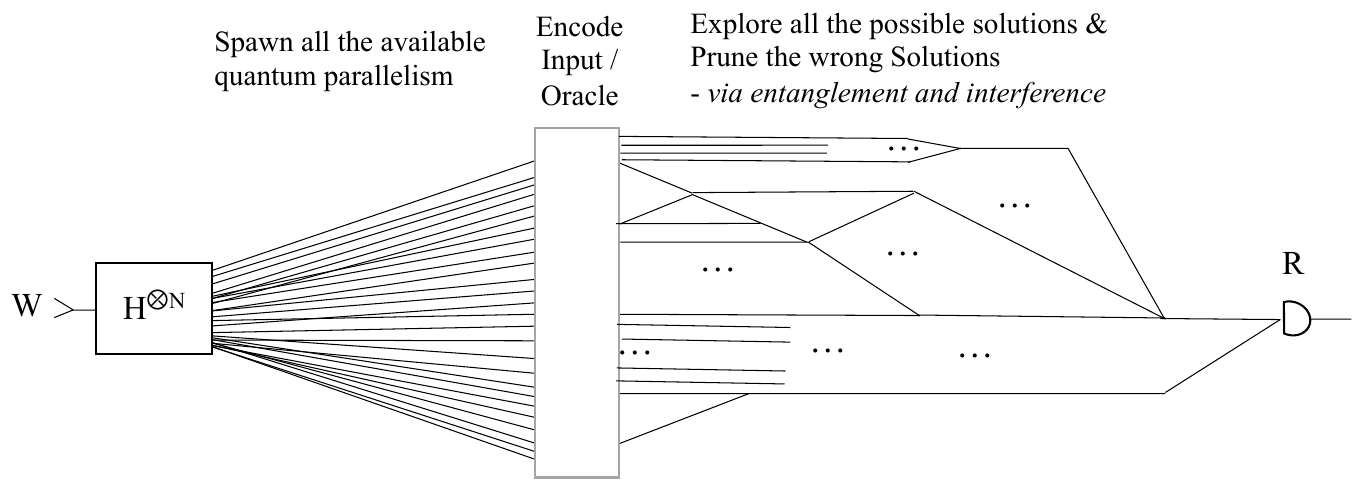}
  \caption{Diagram illustrating a prototypical quantum application workflow. Traditional quantum algorithms typically begin by initializing a classical state, followed by the parallel spawning of quantum parallelism by applying Hadamard gates ($H^{\otimes N}$). Subsequently, the input data is encoded, typically in the quantum state's amplitude and phases, or an oracle is applied. Computational processes then occur in superposition before concluding with a READ operation (measurement). Notably, while the initial phase of the algorithms maximizes quantum parallelism, extracting meaningful results often relies on pruning erroneous outcomes via destructive interference.}
  \label{qapplication}
\end{figure}
When developing quantum algorithms that, by intrinsic nature, are quantum parallel, it is important to rethink how quantum parallelism is used in quantum applications. In fact, quantum algorithms exploit parallelism very differently than classical ones. Classical parallelism is set to do useful work: each thread is set to do an operation that contributes to the overall solution of the problem, e.g., the problem's solution in a certain sub-domain of the overall problem's domain. Instead, quantum algorithms have been set to work differently. A prototypical quantum application is shown in Fig.~\ref{qapplication}.  They exploit Michelangelo's \emph{forza of levare} (means of removal): they spawn maximum parallelism available to explore all the solutions (even the wrong ones) and prune to a small number of threads containing the correct solutions via quantum interference. Quantum algorithms consist of three fundamental phases: \textit{i)} All the available quantum parallelism is spawned out of a classical input (WRITE), typically using parallel execution of \texttt{H} gates, and the input is encoded via an encoding or an oracle is queried exploiting all the available quantum parallelism. \textit{ii)} All the possible solutions are explored via quantum parallelism. The amplitude of correct solutions is amplified via constructive interference, while the amplitudes of incorrect solutions are reduced via destructive interference. Entanglement allows us to explore a larger number of possible solutions that are not expressed by using only single-qubit gate operations. \textit{iii)} A READ or measurement operation is carried out to collapse the superposition to a classical case. Small quantum parallelism before the measurement is a convenient feature for many quantum applications, as it will result in fewer outcomes.

Current Noisy Intermediate-Scale Quantum (NISQ)~\cite{preskill2018quantum} algorithms, such as variational quantum eigensolvers~\cite{peruzzo2014variational, tilly2022variational} and, in general, parametrized quantum circuits~\cite{schuld2015introduction}, do not follow the traditional quantum application pattern, as exemplified in Fig.~\ref{qapplication}. NISQ applications are slightly different as they typically do not include initialization of the application that brings the quantum systems into a full equal superposition of the available quantum parallelism. Instead, the initial superposition is provided by the encoding of the input into the quantum states: phase (or angle) encoding and amplitude encoding have the potential of using all the available superpositions, while basis state encoding provides only limited quantum parallelism.  In particular, the NISQ algorithms lack the initial phase, where the available quantum parallelism is spawned via \texttt{H} gates. Instead, the superposition creation depends on the particular encoding scheme or feature mapping and usage of \texttt{ROTX($\theta$)}  and \texttt{ROTY($\theta$)} quantum gates (and their controlled versions) in the \textit{ansatz} or parametrized quantum circuit~\cite{hidary2019quantum}. Because superposition is connected with the encoding scheme, there is likely a link between quantum parallelism and the expressivity of certain quantum neural network architecture~\cite{schuld2021effect,schuld2021supervised,schuld2021quantum}.

\section{Quantum Parallelism in Classical Parallelism Terms}
If we identify the number of quantum states in superposition in the computational basis as a metric of quantum parallelism, then quantum parallelism can be viewed as a form of \textit{data parallelism}. This analogy arises from the fact that quantum transformations operate simultaneously on the superposition states. Notably, quantum parallelism aligns with the concept of SIMD (Single Instruction Multiple Data) parallelism, where a single instruction operates on multiple processing elements, each handling distinct data elements~\cite{hager2010introduction,sterling2017high}. As discussed previously, traditional quantum application algorithms typically undergo an initial phase where they harness all available parallelism, compute in superposition, manipulate phases, and perform measurements. During this initial phase, the quantum algorithm maximizes parallelism. However, parallelism diminishes as the application progresses due to interference phenomena, reducing the quantum parallelism.

\begin{figure}[t]
 \centering
\includegraphics[width=0.4\linewidth]{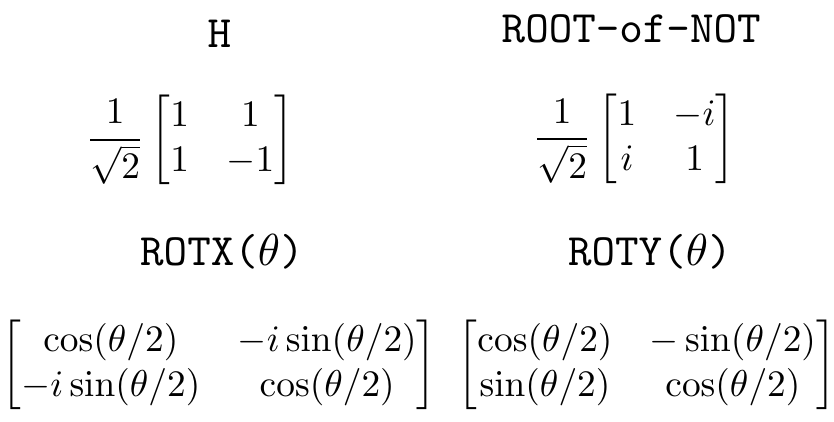}
 \caption{Examples of single-qubit gate operations capable of generating quantum parallelism from a classical state, where no superposition exists. In classical terms, these operations correspond to a fork operation, akin to spawning tasks across multiple processing units.}
\label{matrices}
\end{figure}

In any quantum algorithm, mechanisms for spawning quantum parallelism from a classical input state and mechanisms for reducing data parallelism to identify the correct answer are essential. In classical parallelism, these operations are enabled by fork-join operations. For instance, in classical fork-join paradigms, OpenMP achieves this through opening and closing parallel regions~\cite{haner2022qparallel}, while Cilk utilizes \texttt{spawn} and \texttt{sync} operations~\cite{blumofe1995cilk}. In quantum computing, the fundamental mechanism for creating an equal superposition of basis states, starting from pure basis states, involves using quantum gates with associated dense transformation matrices that yield dense outputs. Examples of such quantum unitary transformations are depicted in Fig.~\ref{matrices}. The Hadamard gate (\texttt{H}), widely regarded as the gateway to quantum parallelism, is the most renowned and ubiquitous gate. Additionally, the previously mentioned \texttt{\footnotesize{ROOT-of-NOT}} gate, as well as the \texttt{ROTX($\theta$)} and \texttt{ROTY($\theta$)} gates, find extensive application in parametrized quantum circuits and variational quantum eigensolvers.

All single-qubit gate operations illustrated in Fig.~\ref{matrices}, along with their associated controlled versions, can generate parallelism from a classical state via quantum interference. The matrices corresponding to these transformations are dense. These operations correspond to a \textit{fork} operation in the classical parallelism context. Due to the unitarity of quantum gates, it is always possible to revert from a superposition of states to a classical state -- transitioning from parallel quantum threads to a single classical thread -- by applying the inverse gates. For example, in the case of \texttt{ROTX($\theta$)} and \texttt{ROTY($\theta$)}, rotating the quantum state by a $-\theta$ angle enables the return to the classical state, thereby reducing the superposition to a single classical state. This reduction in data parallelism via the inverse operation is called the \textit{join} operation. The reduction operation corresponds to a destructive interference phenomenon, resulting in the vanishing probability of a particular path. The \texttt{H} gate is arguably the most prominent bridge for transitioning back and forth between computational basis states and an equal superposition thereof. Although the \texttt{H} gate can be constructed atop \texttt{ROTX($\theta$)} and \texttt{ROTY($\theta$)} gates if not implemented as a standard basic gate, it differs from \texttt{ROTX($\theta$)} and \texttt{ROTY($\theta$)} in that it serves as its inverse. This characteristic grants the \texttt{H} gate a critical role in all quantum algorithms: the same gate can perform fork and join operations.

In this work, we introduce the concept of measuring quantum parallelism using quantum dataflow diagrams, analogous to classical Directed Acyclic Graphs (DAGs) used in parallel computing~\cite{leiserson2008survive}~\footnote{Charles Leiserson, \textit{What the \$\#@! is Parallelism, Anyhow?} \url{https://www.cprogramming.com/parallelism.html}, accessed on May 2024}. Two key metrics quantify classical parallelism: work ($T_W$) and span or critical path length ($T_\infty$). Work, $T_W$, represents the total number of nodes required to execute all operations in the graph, while span, $T_\infty$, represents the longest path of dependencies in the graph in terms of nodes. The classical parallelism $P$ is then calculated as $P=T_W /T _\infty$. To extend this concept to quantum computing, we introduce the quantum dataflow diagram, a graphical tool similar to DAGs but tailored for quantum systems~\cite{vandriessche2013highly}. This diagram represents a weighted graph where vertices correspond to quantum operations, and edges symbolize quantum data movement. By leveraging this representation, we can visualize and quantify quantum parallelism. In the quantum dataflow diagram, each quantum thread is weighted by its probability throughout the execution of different quantum gates. The quantum parallelism quantity enables us to calculate the quantum parallelism efficiency as the fraction of quantum parallelism over the total available parallelism ($\eta_P = P/2^N$), as well as the efficiency of destructive interference in determining correct answers ($\eta_{DI} = 1 - \eta_P$).

\begin{figure}[h!]
\centering
\includegraphics[width=0.6\linewidth]{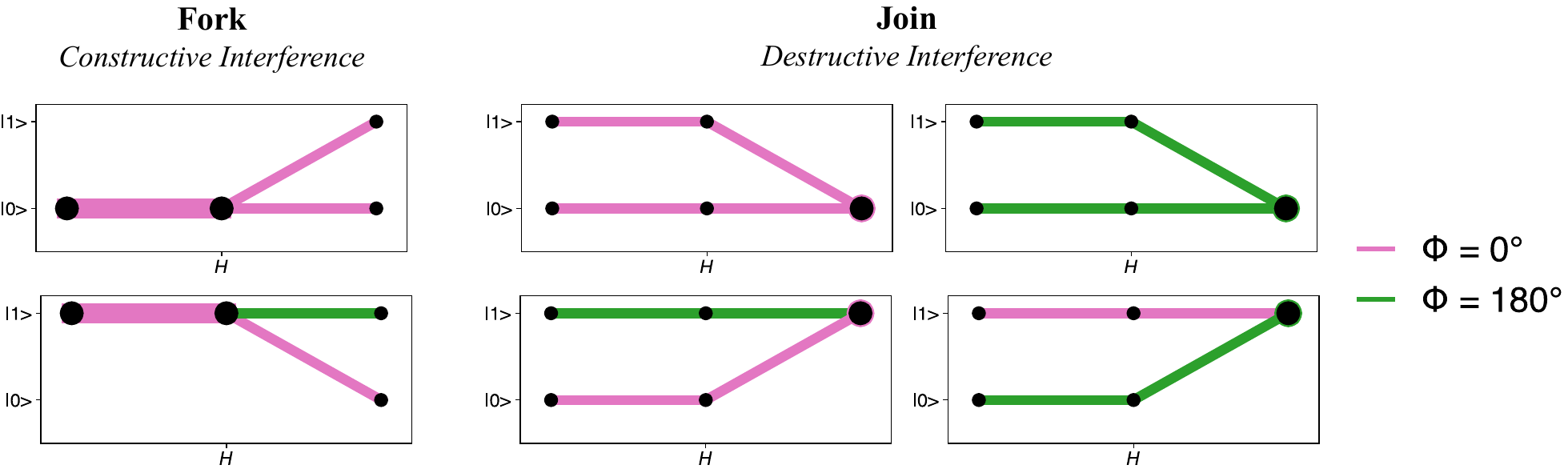}
\caption{The \texttt{H} gate acts as a fork--join operator that allows for creating superposition (fork) and \textit{merging} two quantum states via destructive interference (join). The \texttt{H} gate is the key gate for creating quantum data parallelism with fork-join operations.}
\label{forkjoin}
\end{figure}

To explain the concept of quantum dataflow diagrams, we can focus on expressing the basic fork--join operation obtained with the \texttt{H} gate acting on a qubit. This is shown in Fig.~\ref{forkjoin}.  Note that the \texttt{H} gate exhibits distinct behaviors when applied to the $\ket{0}$ and $\ket{1}$ states, manifesting as a phase difference while sharing identical amplitudes. On the left side of Fig.~\ref{forkjoin}, the\texttt{H} gate can bring a computational basis state, $\ket{0}$ or $\ket{1}$, into an equal superposition of them. As discussed before, differently from the \texttt{ROTX($\theta$)} and \texttt{ROTX($\theta$)}, it is essential to note that the inverse of the \texttt{H} gate is \texttt{H} itself and therefore it can both perform both fork and join operations.

While quantum dataflow diagrams can be manually designed for simple circuits, more complex circuits require simulation tools like \texttt{Qiskit}~\cite{Qiskit} or \texttt{Cirq}~\cite{cirq_developers_2023_8161252,isakov2021simulations} to determine state vectors in polar coordinates, representing amplitudes and phases. The development of automated tools within quantum computing software frameworks can streamline the construction of quantum dataflow diagrams for intricate quantum circuits.

\section{Quantum Dataflow Diagrams Use Cases}
In this section, we apply the methodology of the quantum dataflow diagrams to two important quantum primitives, the QFT and the AA iteration.
\subsection{Quantum Fourier Transform}
\begin{figure*}[t]
  \centering
  \includegraphics[width=\linewidth]{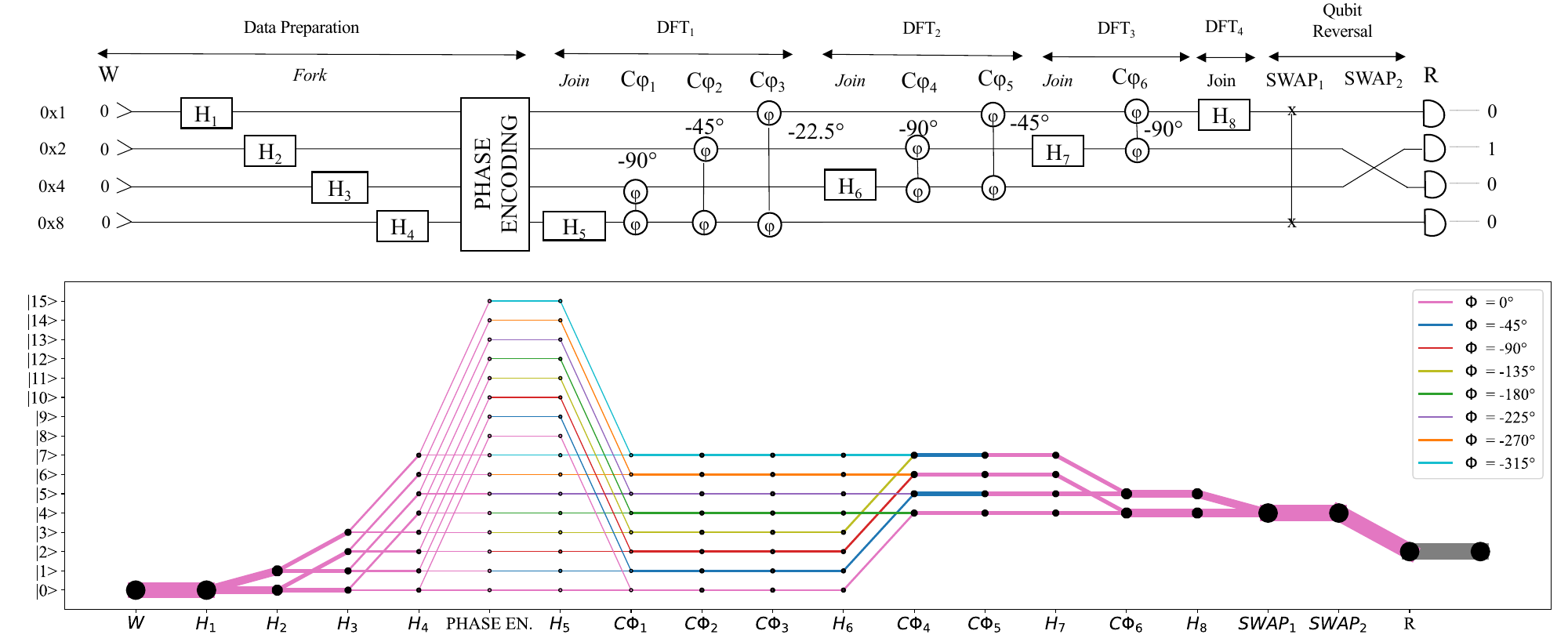}
  \caption{Four-qubit QFT circuit and associated quantum dataflow diagram for the signal of a frequency of two encoded in quantum state phases. }
  \label{QFT}
\end{figure*}
The first example of quantum algorithms we want to analyze for quantum parallelism is the QFT. This method was first devised in 1994 by David Coppersmith~\cite{coppersmith2002approximate}. QFT allows us to calculate the Discrete Fourier Transform (DFT) of a signal encoded in the phase of the quantum states in an equal superposition: it takes the signal included in the quantum state phases and gives as output the frequency of the signal encoded in the amplitude. QFT can also be seen as a powerful way to transform information encoded in the phases into amplitude that can be directly measurable.   Besides the QFT direct usage for spectral analysis, QFT  and its inverse iQFT are omnipresent primitives in quantum computing. They are valuable computing primitives that are critical building blocks for Shor’s algorithms, Quantum Phase Estimation, and Harrow–Hassidim–Lloyd (HHL)~\cite{harrow2009quantum} algorithms, to cite a few prime examples.

 As for the Fast Fourier Transform (FFT), the QFT is based on recursive or iterative execution of smaller DFTs, consisting of controlled phase (\texttt{$C_\phi$}) and \texttt{H} gates~\cite{tucci2004quantum}, and a QFT of one qubit being the \texttt{H}~\cite{cleve2000fast}, with a final qubit reversal, consisting of \texttt{SWAP} gates. When investigating the circuit complexity, the number of gates grows only as  $\mathcal{O}(N^2)$, where $N$ is the number of qubits. Conversely, the classical FFT number of operations grows with the number of input bits $n$ as $\mathcal{O}(n2^n)$. However, the $\mathcal{O}$ notation obfuscates the comparison of FFT and QFT in terms of performance. As noted by several works~\cite{hoefler2023disentangling,johnston2019programming}, there exists a performance crossover point in terms of the number of qubits, after which the QFT becomes more computationally favorable than FFT. In fact, QFT becomes advantageous for more than 22 qubits or about four million input sample sizes. FFT is more convenient than QFT for less than four million input samples. This is one of the reasons why quantum computing is often said to be favorable for solving big-data problems.

When designing a quantum dataflow diagram for the QFT, for clarity of exposition, we choose a four-qubit system and associate quantum circuit, as shown in Fig.~\ref{QFT}. We encode a signal of frequency two in the phase of the quantum states: the phase rotates twice across quantum states, and the result is deterministic and equal to $\ket{2}$, when measured. More precisely, this signal encodes 16 values with relative phases corresponding to two full anti-clockwise rotations (0\textdegree, -45\textdegree, -90\textdegree, ... ) into the quantum states $\ket{0} ... \ket{15}$.  We assume that the encoding operation is only one operation while the encoding phase (such as a simple signal can be encoded by using a quantum circuit).  After putting the system in a superposition of states and encoding the input vector with a quantum circuit, four quantum DFTs are applied in succession, with a final qubit reversal of the quantum results using \texttt{SWAP} operations. 

When analyzing the QFT dataflow diagram, we have a serial WRITE operation during the data preparation, which is the interface between classical and quantum worlds. We then have four Hadamard gates (\texttt{$H_1$} -- \texttt{$H_4$})  to put in a superposition using a fork operation. \texttt{$H_1$} -- \texttt{$H_4$} spawn the maximum available parallelism, which is 16 quantum threads. The QFT algorithm consists of four quantum DFT blocks with \texttt{H} gates (the \texttt{H} gate itself can be considered as a DFT of size one), providing destructive interference and controlled phase operations that strategically alter the phases to allow for more interference. In particular, the Hadamard gates \texttt{$H_5$} -- \texttt{$H_8$} halve the quantum parallelism, similarly to the \textit{decimation} transformation in FFTs. After \texttt{$H_8$}, the process is inherently serial (or classical) with the qubit reversal operation via $\texttt{SWAP}$ gates,  required to provide the correct results.  We want to point out, though, that the final \texttt{SWAP} gates are critical to providing entanglement, necessary entangled states that could not be provided by other means. For instance, consider a signal that consists of a superposition of constant (equivalent to frequency 0) value and a frequency equal to 15: the QFT output is an entangled state, only achievable via the final \texttt{SWAP} gates.  

It is important to acknowledge that varying the input leads to changes in the amount of parallelism. For instance, an input signal featuring multiple frequencies results in higher parallelism and diminished destructive interference. The quantum parallelism is intimately tied to the input and encoding methodologies employed. 

An intriguing observation arises from the reversibility of quantum transformations: the quantum dataflow diagram for the inverse QFT (iQFT) can be interpreted in the reverse direction. For instance, one can trace the path from a frequency of two to create a signal encoded in the phase of quantum states in equal superposition.


\begin{figure}[t]
\centering
\includegraphics[width=0.6\linewidth]{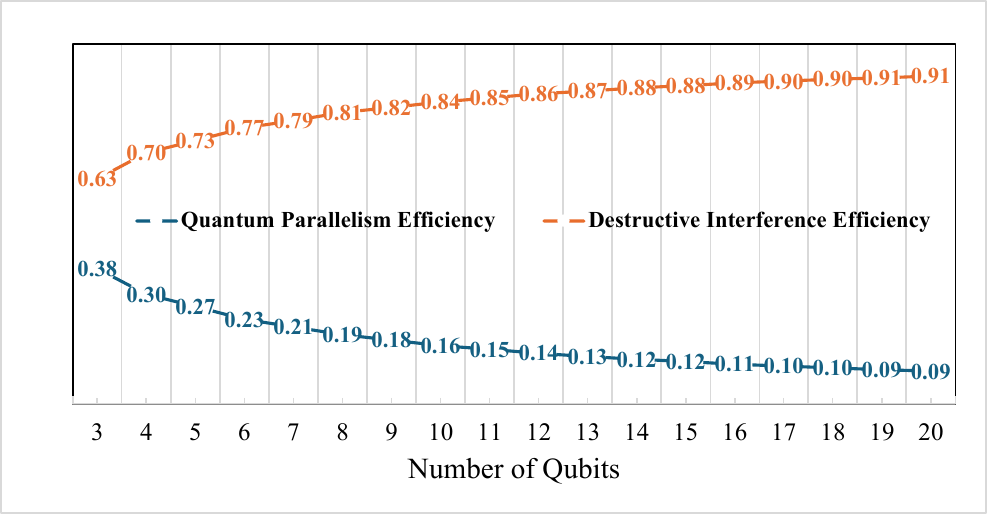}
\caption{Quantum parallel and destructive interference efficiencies varying the number of qubits in a QFT acting on a signal with a frequency of two. }
\label{efficiency}
\end{figure}
\begin{figure*}[h!]
\centering
\includegraphics[width=\linewidth]{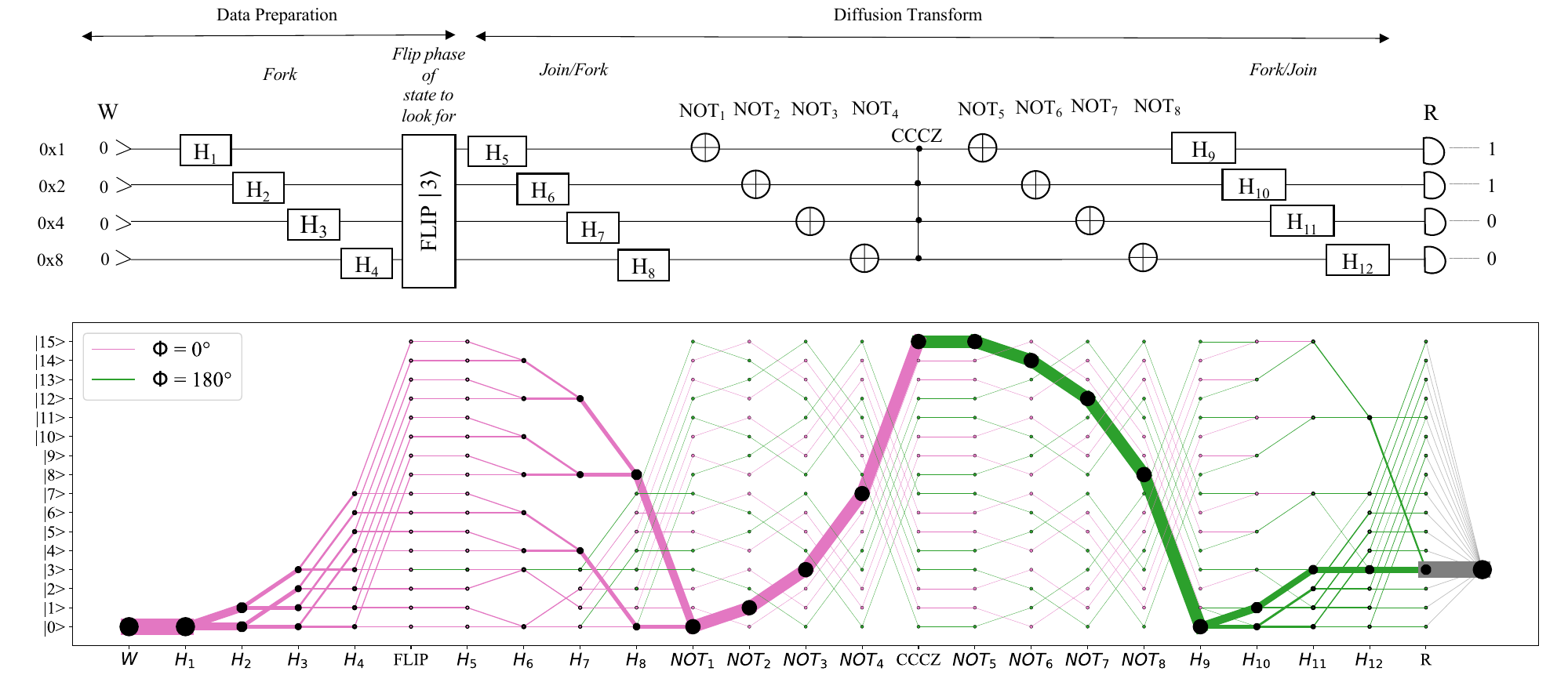}
\caption{Quantum dataflow diagram for one AA iteration in a four-qubit system. In this example, the flip operation marks the relative phase of the quantum state $\ket{3}$.}
\label{AAiteration}
\end{figure*}

The QFT dataflow diagram provides insights into the quantum parallelism inherent in the algorithm. Simplifying the analysis to focus on the parallelism after data preparation, including the serial W operation and the initial encoding of the input signal, we can calculate the total number of nodes ($T_W = 68$) and the critical path length ($T_\infty = 14$). This results in a total parallelism of $P = T_W / T_\infty = 4.86$. With corresponding values of $\eta_p = 0.3$ and $\eta_{DI} = 0.7$, we observe that the quantum parallelism accounts for only 30\% of the available parallelism of 16, indicating effective destructive interference. By fixing the QFT input signal to a frequency of two, we can derive an expression to calculate the quantum parallelism for varying numbers of qubits as follows: $P  = (N 2^N + \lfloor N/2 \rfloor + 2 ) / ( \frac{1}{2} N (N+1) + \lfloor N/2 \rfloor + 2)$. As the number of qubits increases, the quantum parallelism efficiency decreases while the efficiency of the destructive interference increases. This trend is depicted in Fig.~\ref{efficiency}.

\subsection{Amplitude Amplification Iteration}
The AA iteration (also called \textit{diffusion transform} or \textit{inversion above the average}) is a critical quantum computing primitive, first designed and expressed in the formulation we know today by Lev Grover in 1996~\cite{grover1996fast,grover1997quantum,grover1998quantum}. Lev Grover applied AA iterations to the problem of finding an item in an unordered set, the so-called \textit{searching for a needle in a haystack} problem~\cite{grover1997quantum} and pointed out that the basic AA iteration can also be applied to several other applications, including neural networks and associative memories. With $N_I$ items, Grover's algorithm relies on $\sqrt{N_I}$ AA iterations (including the query/ call to the oracle) to achieve the maximum probability of measuring the correct item. In the classical case, identifying an item would take $N_I -1$ queries in the worst-case scenario or $N_I/2$ queries on the average. For this reason, Grover's algorithm provides only a quadratic speed-up, when compared to the classical case.

At its fundamentals, AA transforms the difference between the relative phases into the amplitude magnitude, making the quantum state that, for instance, we marked with a flipped relative phase, identifiable through a measurement. For the sake of simplicity, we flip the phase of a given quantum state by 180\textdegree. However, this can be implemented with an oracle based on phase logic. 

Fig.~\ref{AAiteration} illustrates the quantum circuit for the AA iteration on a four-qubit system and marks the relative phase of the quantum state $\ket{3}$ by flipping its phase. The circuit consists of data preparation, including spawning all available parallelism and encoding the input as a flipped relative phase in the quantum state, and the diffusion transform quantum circuits. The outcome of the AA quantum primitive transforms the relative phase difference into an amplitude difference, enhancing the probability of measuring state $\ket{3}$. 

When investigating the AA quantum dataflow diagram, we observe that, as for the QFT, the write operation is serial. After the data preparation step (corresponding to all the available parallelism and marking an item by flipping its corresponding quantum state phase),  and the \texttt{$H_5$} -- \texttt{$H_8$} transformations,  AA exhibits a quantum thread with a larger amplitude (corresponding to thicker lines) than others.  We might be tempted to prune the quantum dataflow diagram to one quantum thread, de-quantize it, and make it serial to all the quantum states resulting from the sequence \texttt{$NOT_1$} -- \texttt{$NOT_8$}. However, this would not be the correct method to use, and it would break the correctness of the algorithms. The AA relies critically on all the 16 quantum threads -- albeit they have a low probability of occurring -- to achieve the correct answer. To convince ourselves of that, hypothetically, we can merge all the quantum threads into the dominant one between \texttt{$NOT_1$} and \texttt{$NOT_8$}. However,  \texttt{$H_9$} -- \texttt{$H_{12}$} acting only one state quantum will create an equal superposition of states with no discernible result when measuring. In other words, the unbalanced superposition across all the 16 quantum states is critical for the correct code functioning, and therefore, we cannot prune the different threads. We note that the AA iteration and Grover's algorithm use all the available parallelism by design, resulting in a non-deterministic outcome.

As it was done for the four-qubit QFT, we proceed to evaluate the quantum parallelism inherent in the AA iteration. Disregarding the data preparation phase, in the context of a four-qubit system, we observe a total of 242 nodes ($T_W$) and a span of 19 ($T_\infty$), resulting in a quantum parallelism of $P=T_W /T \infty = 242/19 = 12.7$. A quantum parallel efficiency of $\eta_P =  12.7 / 16 = 0.8$ indicates that the AA iteration uses 80\% of the available parallelism, showcasing a higher utilization of available quantum parallelism when compared to the QFT. Moreover, the influence of destructive interference with a $\eta_{DI} = 0.2$ appears to be relatively limited in the AA iteration scenario. As for the QFT, the efficiency of quantum parallelism diminishes with an increasing number of qubits, while the efficiency of destructive interference shows a proportional increase.

\section{Quantum Parallelism Laws}
An intriguing consideration in the context of quantum computing is the applicability and formulation of classical parallelism laws, such as Amdahl's and Gustafson's laws, in quantum computing. Amdahl's law states that the overall performance improvement gained by optimizing a single part of a system is limited by the fraction of time of the serial~\cite{amdahl1967validity}. In other words, the maximum speed-up achievable (i.e., overall performance improvement) depends on the percentage of the code that can be parallelized, regardless of the amount of available parallelism. In quantum algorithms, parallelism is an inherent characteristic associated with the quantum circuit and depends on both the algorithm and the input data.

\subsection{Reconsidering Amdahl's Law for Quantum Computing}
The Amdahl's law describes the potential speed-up of a classical parallel computing system as a function of the proportion of the program that can be parallelized, and it can be expressed as follows:
\begin{equation}
S = \frac{1}{F + (1 - F)/P},
\end{equation}
where $S$ is the parallel speed-up, $F$ is the fraction of the threads/processes that must be run serially (that cannot be parallelized), and $P$ is the number of threads/processes. In the limit of infinite parallelism ($P  \rightarrow  \infty $), the speed-up of a program with a finite serial part is asymptotically equal to $1/F$. For instance, if the serial portion of a program constitutes 50\% of the whole program, even with an infinite number of processes, the speed-up is only two. 

While Amdahl's law provides valuable insights into the limitations of parallelism in classical systems, its direct application to quantum computing faces challenges due to the unique characteristics of quantum algorithms. In quantum computing, parallelism is an inherent characteristic associated with the quantum circuit, and it depends on both the algorithm and the input data. However, the speed-up achieved does not directly depend on quantum parallelism and may even decrease with it. This is because increased quantum parallelism often leads to less efficient interference patterns, limiting the effectiveness of parallelism. Additionally, Amdahl's law is formulated with strong scaling in mind, where the problem size remains constant, and parallelism is increased. However, in quantum computing, increasing parallelism automatically increases the workload as more qubits are used. Therefore, the concept of strong scaling does not directly apply to quantum computing.


While we argued that the speed-up does not depend on quantum parallelism, the fundamental limitation related to the serial portion of the algorithm still holds, e.g., a serial portion in a parallel algorithm can nullify computing mechanisms to achieve speed-up. All the quantum algorithms have an inherently serial part, which is the classical quantum I/O, and in general, quantum data preparation and retrieval are needed to initialize the quantum simulation~\cite{hoefler2023disentangling}. For instance, when solving a linear system with the HHL algorithm~\cite{harrow2009quantum}, the matrix must be loaded from classical data and encoded into the quantum state. The classical-quantum I/O wall fundamentally limits the quantum advantage, regardless of the speed-up achieved with quantum computing. For this reason, as pointed out by several previous works, the serial fraction due to the I/O wall has to be tackled as one of the priorities. Another more fundamental factor limiting the quantum advantage is the lack of algorithm destructive interference, which is key to eliminating the wrong, incorrect solutions. 

\subsection{Reconsidering Gustafson's Law for Quantum Computing}
Conversely to Amdahl's law, Gustafson's law~\cite{gustafson1988reevaluating} focuses on scaling the size of the problem as the number of processors increases. It assumes that the workload and problem size can be scaled up to use additional processors in a setup called \textit{weak scaling}. Gustafson's law expresses the speed-up as $S = P - F (P -1)$. In the classical context, Gustafson's law effectively removes the limitation of the serial part, making the speed-up close to linear with the number of processors if the non-parallelizable fraction is relatively small. In quantum computing, the inherent serial part is the classical quantum I/O, which is not a constant factor, but instead, it is a function of the number of quantum states used for encoding the input. Performance models for describing the classical-quantum I/O wall are needed further to develop Gustafson's law to the quantum regime.

Another interesting point related to Gustafson's law and weak scaling is the quite narrow weak scaling window in terms of qubits, where quantum computing is effectively advantageous for classical computing. For relatively small problem sizes, classical computing is faster. We saw that we need at least an input size of $2^{22}$ for the QFT to make it advantageous. In addition, given the exponential growth of quantum states with the number of qubits, quantum computing indeed provides an unparalleled number of superpositions of quantum states, and the available quantum parallelism (albeit not the quantum parallelism efficiency) increases exponentially with the number of qubits. This presents a fundamental departure from the classical case, where achieving significant advances in parallelism requires substantial technological leaps, and still, scientific workloads can be tailored to leverage the available parallelism efficiently. In quantum computing, neglecting the problem of the factorization of very large numbers, we find ourselves in a new regime where the amount of parallelism surpasses the needs of even the largest problem sizes. In such a paradigm, we witness the end of Gustafson's law, as the problem size can no longer effectively utilize the potential of potentially unlimited parallelism offered by quantum computing because of \textit{limited classical} problem sizes.

\section{Conclusion} 
This work investigated the fundamental questions surrounding quantum parallelism in quantum applications, exploring its implications for actual speed-up and challenging classical parallelism paradigms in the context of quantum computing. Quantum parallelism, emerging from the interaction of $2^N$ quantum states, exhibits a characteristic interference pattern reminiscent of waves or antennas. By leveraging classical parallel computing concepts within quantum computing, we underscored the importance of fork-join operations, which enable both constructive and destructive interference. To quantify parallelism, we introduced quantum dataflow diagrams, providing a visual tool for measuring quantum parallelism in practical applications such as the QFT and AA iterations. Additionally, we introduced metrics such as quantum parallelism and destructive interference efficiencies to evaluate the utilization of available parallelism in quantum algorithms. Our study revealed that in the QFT and AA iterations, quantum parallel efficiency tends to diminish with an increased number of qubits for the analyzed applications. Finally, drawing parallels with classical computing principles, we revisited Amdahl's and Gustafson's laws. This led us to reconsider the role of the serial portion of quantum applications, the classical-quantum I/O bottleneck, and highlighted the significance of a narrow window of problem sizes that render quantum computing advantageous.

An inherent challenge in defining quantum parallelism lies in its entanglement -- pun intended -- with the interpretation of quantum mechanics, which leads to philosophical complexities regarding the actual realization of quantum parallelism and computation~\cite{duwell2007many, duwell2018make, lanzagorta2008quantum}. Different interpretations of the physical reality of quantum mechanics can give rise to distinct \emph{conceptual} implementations of parallelism. To navigate this challenge, we adopted a practical yet shallow approach, circumventing philosophical obstacles as much as possible. However, it is worth noting that one of the most intuitive and elegant interpretations is Hugh Everett's many-worlds interpretation of quantum mechanics and the multiverse hypothesis~\cite{wallace2012emergent,deutsch1998fabric}. According to this interpretation, time is envisioned as a many-branched tree, where every possible outcome of quantum parallelism is actualized in a separate branch or universe. This interpretation suggests that each computational path exists simultaneously across different branches of reality. This concept aligns with the notion of quantum parallelism, implying that all potential outcomes of quantum computation occur in parallel across multiple universes. One important implication of the multiverse theory is its ability to support quantum parallelism surpassing the number of particles in the observable universe (\textgreater 300 qubits). In such scenarios, multiple parallel universes can concurrently accommodate computational processes unfolding across different branches without being constrained by the particle count within a single universe~\cite{deutsch1998fabric}.

This work offers an intuitive perspective on quantum parallelism, clarifying its fundamental aspects. Nevertheless, it is crucial to acknowledge that our exploration merely scratches the surface of quantum parallelism. Our investigation primarily revolves around quantum parallelism within the computational or standard basis, aligning with the canonical approach directly corresponding to classical bit concepts. However, using alternative bases would necessitate a reconfiguration of the quantum dataflow diagrams. Moreover, we focused on quantum gate abstractions, state vectors, and pure state formulations. Alternative and more comprehensive quantum computing abstractions and approaches, such as Hamiltonian parameterization~\cite{peng2024simuq} or the exploration of continuous-variable quantum computing paradigms~\cite{lloyd1999quantum,killoran2019continuous}, require further investigations. Such approaches could offer even deeper insights into quantum parallelism. 

From the practical point of view, to quantify the extent of quantum parallelism, we proposed the quantum dataflow diagram, akin to classical dataflow graphs~\cite{wongsuphasawat2017visualizing}. This tool enables the visualization and measurement of quantum parallelism, revealing critical distinctions between classical and quantum parallelism and facilitating the development of quantum performance models. For simplicity, we have assumed a uniform work associated with each gate transformation, where each node has an execution time of one. However, quantum dataflow diagrams can be used as starting models to develop quantum application performance models. A more accurate model of the quantum application can be achieved by weighting each node differently based on its computational cost. Moreover, we can incorporate additional overheads, such as latencies from quantum hardware, into the dataflow diagram to construct more sophisticated quantum application performance models.

This paper aimed to pave the way for expressing quantum concepts within the framework of classical parallelism. While this approach holds promise for practical applications in quantum programming models—especially those built upon classical programming paradigms and tools—it remains uncertain whether it can be directly applied to specific hardware concepts. This uncertainty arises because the paper primarily deals with high-level abstractions, such as quantum gates and circuits, which are agnostic to specific hardware implementations. Nevertheless, this work offers a perspective that can stimulate further research and engage the HPC community on the elusive concept of quantum parallelism. By bridging classical parallelism frameworks with quantum computing principles, the paper opens doors for exploring methodologies, developing quantum programming models, and advancing our understanding of quantum algorithms.

\end{document}